\documentclass[12pt]{article}
\usepackage[english]{babel}
\usepackage{times}
\usepackage[T1]{fontenc}
\usepackage{epsfig}
\usepackage{amsmath}
\usepackage{amssymb}

\begin{document}

\title{Relativistic quantum mechanics of  the Majorana particle:  quaternions, paired plane waves, and orthogonal representations of the  Poincar\'e group }

\author{H. Arod\'z$\:^1$ $\;$ and  $\;$  Z. \'Swierczy\'nski$\:^2$ \\ 
	{\small $^1$
Jagiellonian University, Cracow, Poland} \footnote{henryk.arodz@uj.edu.pl} \\ {\small $^2$ Pedagogical University, Cracow, Poland} \footnote{zs@up.edu.pl} }
\date{$\;$}

\maketitle

\vspace*{0.5cm}

\begin{abstract}
 The  standard momentum operator $-i \nabla$  can not be accepted as observable in relativistic quantum mechanics of the Majorana particle.   Instead, one can use  axial momentum operator recently proposed in Phys. Lett. A {\bf383},  1242 (2019).   In the present paper we report several  new results related to the axial momentum which elucidate its usability.  First, a new motivation  for the axial momentum  is given, and the Heisenberg  uncertainty relation  checked. Next, we show that the general solution of time evolution equation   in the axial momentum basis has a connection with quaternions.  Single traveling  plane waves  are not possible  in the massive case, but  there exist solutions  which consist of  asymmetric pair of  plane waves traveling in  opposite directions. Finally, pertinent real orthogonal and irreducible representation of the Poincar\'e group --  consistent with the lack of antiparticle -- is unveiled. 
\end{abstract}

\vspace*{0.9cm}

\pagebreak

\section{ Introduction}
The discovery  of non vanishing mass of neutrinos has led to many conjectures about the nature of these particles.  In particular,  it  is possible that they are  relativistic massive fermions of the Majorana type.  While  state of the art description of fundamental particles is provided by quantum field theory  (notwithstanding its well-known problems),  the slightly older framework of relativistic quantum mechanics  also is useful,  especially in the case of  single particle.  
Relativistic quantum mechanics  has  many important  applications in such branches of contemporary physics  as atomic physics, theory of elementary particles, and even condensed matter physics, see, e.g.,   \cite{1}, \cite{2} and \cite{3}. Among  relativistic wave equations, the most popular is of course the one proposed by P. A. M.  Dirac, but other equations are interesting  as well,  in particular  the Proca   and the  Salpeter equations recently discussed in, respectively,  \cite{4} and \cite{5}.  There is no doubt that  relativistic quantum mechanics is the source of important insights.  Quantum mechanics of the Majorana particle is not  exception to this.

Relativistic  quantum mechanics of a free Majorana particle  significantly differs from quantum mechanics of the Dirac particle,  and  it has several unexpected features.  Certain aspects of it  have already been considered in \cite{6}, \cite{7},  \cite{8},  \cite{9} and \cite{10}. There is an interesting problem concerning  momentum observable, to the best of our knowledge first considered in \cite{7}, and  recently readdressed  in \cite{10} where the axial momentum  operator has been proposed as the momentum observable.     In the present paper  we  apply  expansions into eigenfunctions of the axial momentum operator in order to discuss general solution of the  time evolution equation  as well as the relativistic invariance.

The time  evolution equation for the Majorana particle coincides with the Dirac equation \footnote{We use the natural units, $c = \hbar =1$.},
 \begin{equation}
 i \gamma^{\mu} \partial_{\mu} \psi - m \psi =0,
 \end{equation}
  in which  certain  Majorana-type  representation 
 for  Dirac matrices $\gamma^{\mu}$ is assumed, that is  all these matrices are  purely imaginary. The crucial difference with the Dirac particle is that  all four components of the  bispinor $\psi$   are real numbers. This is consistent with Eq.\ (1) because the matrices $\gamma^{\mu}$ are imaginary, and $m $ is a non-negative real number. In consequence, the relevant Hilbert space  ${\cal H}$ consists of  all real normalizable bispinors, and the pertinent algebraic number field  is that of real numbers $\mathbb{R}$, instead of the more common 
in quantum mechanics algebraic field of complex numbers $\mathbb{C}$. 
Let us note  that there are other formulations of  Majorana quantum mechanics,  see, e.g.,  \cite{6} and \cite{9},  but they are  equivalent to the one adopted  here.

Quantum mechanics with real numbers or even quaternions in place of the algebraic field of complex numbers is not very popular, but it has been thoroughly discussed in literature, see for example \cite{11}, \cite{12} and \cite{13}. In particular, it is known that in the real and quaternionic cases the discrete symmetries  $P$, $T$, and $C$ are represented by unitary operators, while in the complex quantum mechanics also anti-unitary symmetry operators can appear.

The  standard momentum operator $\hat{\mathbf{p}}= - i \nabla$  turns  real bispinors  into  imaginary ones,  therefore it is not  operator in ${\cal H}$.  New momentum-like operator  is needed.   Such operator -- called  axial momentum and denoted  $\hat{\mathbf{p}}_5$ -- has been proposed  in \cite{10}, namely \[ \hat{\mathbf{p}}_5 = -i \gamma_5 \nabla  \] 
in the Schroedinger picture.  It is Hermitean, and its  spectrum is continuous.  Moreover,  it can be regarded as the generator of spatial translations. On the other hand, it has  certain rather peculiar features.  First,  it does not commute with Hamiltonian   in the case of massive Majorana particle   ($m >0$) . In consequence,  its direction  is not constant in time in the Heisenberg picture --   the axial momentum contains a rotating component of the magnitude $m/E_p$, where $E_p=\sqrt{m^2 +\mathbf{p}_5^2} $ is the energy of the particle \cite{10}. This component is  negligibly small at high energies, but  it can not be neglected at the energies comparable with $m$.   The eigenfunctions of the axial momentum 
can be used as a basis for Fourier-type expansion of  time-dependent wave functions of the Majorana particle \cite{10}. It turns out that in place of simple  time-dependent  $U(1)$  phase factors  known from the case of  Dirac particle  there are cumbersome time-dependent $SO(4)$  matrices.

In the present paper we continue the investigations  initiated in \cite{10}.    We begin with a new motivation for the axial momentum operator. It  is  based on a mapping between  the Majorana and Weyl bispinors.   We show that  the classic position vs. momentum uncertainty relation   remains unchanged when the standard momentum operator is replaced by the axial one.  Next, we examine  the  general solution of the wave equation  expanded in the basis of eigenfunctions of the axial momentum  $\psi_{\mathbf{p}}(\mathbf{x})$.   It turns out that  it can be regarded as  position and time dependent  quaternion.  Next, 
 we transform the solution  to a more convenient form which does not contain the $SO(4)$ matrices, namely we rewrite it as a superposition of traveling plane waves, see formula (13)  below.   That this is at all  possible   is a surprise because the direction of the axial momentum is not constant in time if $m>0$.   Interestingly, it turns out that in the massive case the plane waves necessarily come in pairs. The paired  plane waves have  the opposite  wave vectors $\mathbf{p}$ and   $- \mathbf{p}$, hence they travel in the opposite directions.  Their amplitudes  are not equal,  the ratio is  $1: m/E_q $.

Finally, we elaborate on the Poincar\'e  invariance of the model  using  the   amplitudes introduced in the basis of  eigenfunctions of the  axial momentum. We find rather straightforward  realization of the relativistic invariance in the space of these amplitudes, which is rather  encouraging result. The obtained  representation of the Poincar\'e group   in the massive case is orthogonal,   irreducible, and equivalent to real version of the well-known  spin 1/2  unitary irreducible  representation. Recall that in the case of Dirac particle one obtains a  reducible representation composed of  two  spin 1/2  irreducible representations.  
 Such representations discussed  in the framework of relativistic quantum mechanics  usually  reappear unchanged when one considers single particle sectors  for the related  quantum field.  In the Dirac case the two spin 1/2 representations correspond to the particle and its antiparticle. In the Majorana case we expect the particle only.  

Our overall conclusion is that   the axial momentum is  a reasonable replacement for the ordinary momentum  (which should not be used in the Majorana case anyway).  Certain peculiar features present in the case of massive Majorana particle, like the discussed in Section 3 mixing of modes with  opposite  axial momenta,   are  negligible at  energies  $E_p \gg m$.  At lower energies  however they can not be neglected. We regard them as  intrinsic physical features of the relativistic massive Majorana particle.

The paper is organized as follows. In the next section  we  introduce the axial momentum  using the mapping between the Weyl and Majorana bispinors and we derive the uncertainty relation.   In Section 3, after a brief  recap of necessary results from  \cite{10},  we point out the connection with quaternions, and we discuss the traveling plane waves. Section 4 is devoted to analysis of  the representation of the Poincar\'e group  that exists in the space of solutions of the evolution equation.

\section{The axial momentum:  new motivation, and  the Heisenberg uncertainty relation}
Throughout this paper we work with the Dirac matrices $\gamma^{\mu}$ in a  Majorana-type representation, i.e., the matrices are purely imaginary. Then also the matrix $\gamma_5 = i \gamma^0 \gamma^1 \gamma^2 \gamma^3$ is purely imaginary. Furthermore,  $\gamma_5$ is Hermitian,  hence also anti-symmetric:   $\gamma_5^T = - \gamma_5$, and  $\gamma_5^2 = I$ , where $I$ is the 4 by 4 unit matrix.  We work with the following set of the Dirac matrices
\[ \gamma^0 = \left(  \begin{array}{cc} 0 & \sigma_2 \\ \sigma_2 & 0 \end{array}  \right), \;\;    \gamma^1 = i \left(  \begin{array}{cc} - \sigma_0 & 0 \\ 0 & \sigma_0 \end{array}  \right), \;\;   \gamma^2 = i \left(  \begin{array}{cc} 0 & \sigma_1 \\ \sigma_1 & 0 \end{array}  \right),\]  \[  \gamma^3 = -i \left(  \begin{array}{cc} 0 & \sigma_3 \\ \sigma_3 & 0 \end{array}  \right),  \;\;\;\;\; \mbox{and} \;\;\;\;\;  \gamma_5 = i \left(  \begin{array}{cc} 0 & \sigma_0 \\ - \sigma_0 & 0 \end{array}  \right).  \]
Here $\sigma_k$ are the Pauli matrices, and $\sigma_0$ is the 2 by 2 unit matrix.

In all Majorana-type representations  charge conjugation $C$ is represented   just by the complex conjugation.  Therefore, the Majorana bispinors, which by definition are  invariant  under $C$, have only real  components in such representations. The operator  $\hat{\mathbf{p}}_5$  commutes with $C$, in contradistinction to  $\hat{\mathbf{p}}$.

The motivation for the axial momentum given in \cite{10} refers to a Lagrangian in  classical field theory and to the Noether theorem.  Moreover,  it applies to the massless case ($m=0$) only.  Below we give an independent, simple and  more general motivation.

 There is a simple one-to-one mapping $M$ between linear spaces of the Majorana  bispinors and  right-handed (or left-handed) Weyl bispinors.   From 
arbitrary right-handed Weyl bispinor $\phi$, which by definition has the  property $ \gamma_5 \phi = \phi$, we form $ \psi =  \phi + \phi^* \equiv M(\phi)$, which is real, hence  Majorana, bispinor. The asterisk denotes the complex conjugation.  Now, because  the matrix $\gamma_5$ is purely imaginary,   $\phi^*$  is a left-handed bispinor, $\gamma_5 \phi^*= - \phi^*$.  It follows that  $ \gamma_5 \psi =   \phi - \phi^*$,  and  $\phi = (I + \gamma_5)\psi/2 \equiv \psi_{R}$,  $\phi ^*= (I - \gamma_5)\psi/2 \equiv \psi_{L}$.   This shows that the mapping $M$  is invertible. Notice that  it preserves linear combinations  only if their coefficients are real.
  The Weyl bispinors are complex, hence  the standard momentum operator $\hat{\mathbf{p}} = - i \nabla$ is well-defined for them. In particular, it  commutes with the $\gamma_5$ matrix, therefore also  $\hat{\mathbf{p}}\phi$   is  right-handed Weyl  bispinor.
  Let us find the Majorana bispinor that corresponds to $\hat{\mathbf{p}}\phi$:   \[ M( \hat{\mathbf{p}}\phi)= \hat{\mathbf{p}} \phi  + (\hat{\mathbf{p}}\phi)^* = \hat{\mathbf{p}}( \psi_R -  \psi_L) =  -i \gamma_5 \nabla \psi  = \hat{\mathbf{p}}_5 \psi.\] 
We see that the standard momentum operator in the  space of right-handed Weyl bispinors gives rise to the axial momentum operator in the space of Majorana bispinors. 

 The axial momentum  commutes with $\gamma_5$, therefore it can be used also in the space of right-handed Weyl bispinors. However,  in this space it coincides with $\hat{\mathbf{p}}$ because  $\gamma_5\phi= \phi$.     

 The presence of the  one-to-one mapping $M$ might suggest that  the two  quantum mechanics, Majorana and Weyl,   are equivalent to each other.   For the equivalence,  the mapping $M$ should preserve scalar product. It turns out that it is not the case. Let us take $\psi_1= M(\phi_1), \; \psi_2 = M(\phi_2)$,  and compare  the  scalar product of the Majorana bispinors  $\int\! d^3x \: \psi^T_1 \psi_2$ with the scalar product  $\int\! d^3x \: \phi^{\dagger}_1 \phi_2$ of the corresponding Weyl bispinors. We have
\[  \int\!\! d^3x \: \psi_1^T \psi_2   = \int\! \!d^3x\:(\phi_1^T + \phi_1^{\dagger}) (\phi_2 +\phi_2^*) = \int\!\! d^3x\:(\phi_1^{\dagger} \phi_2 + (\phi_1^{\dagger} \phi_2)^*).  \]
Here we have used the identity $\phi_1^T \phi_2 \equiv 0$,  which follows from the antisymmetry of $\gamma_5$: $\;\;\;\phi_1^T \phi_2 = \phi_1^T \gamma_5\phi_2=- \phi_1^T \gamma_5^T\phi_2= -(\gamma_5\phi_1)^T \phi_2 = - \phi_1^T \phi_2.$
Thus we see that in general the scalar product is not preserved  by $M$, 
  \[ \int\!\! d^3x \: \psi_1^T \psi_2  \neq \int\!\! d^3x \: \phi_1^{\dagger} \phi_2.   \] 
Note also the differences in evolution equations. In the Weyl case,  the evolution equation has the form (1) with $m=0$, namely $  i \gamma^{\mu} \partial_{\mu} \phi =0$, while in the Majorana case $m\neq 0$ is allowed.  Using the mapping inverse to $M$ one can  of course transform Eq.\
(1) for the Majorana bispinor $\psi$ to the space of right-handed Weyl bispinors -- we obtain  
$  i \gamma^{\mu} \partial_{\mu} \phi - m \phi^* =0,$
which is known as  the Dirac equation for $\phi$  with the Majorana mass term (recall that $\phi^*$ is the charge conjugation of $\phi$).   
This last equation can not be accepted as a quantum mechanical evolution equation for the Weyl bispinor $\phi$ because  it is not linear over $\mathbb{C}$ -- it is linear only over $\mathbb{R}$.  The point is that the Hilbert space of the right-handed Weyl bispinors \footnote{By definition, it includes all bispinors which obey the condition $\gamma_5 \phi = \phi$. Arbitrary complex linear combination  of such bispinors  also is right-handed.  The complex conjugate bispinor $\phi^*$ is left-handed.} is linear over $\mathbb{C}$,  therefore also quantum mechanical evolution equation for these bispinors  should be linear over $\mathbb{C}$, otherwise  the superposition principle  is broken.  

Commutator of the axial momentum  with position operator $\hat{\mathbf{x}}$  has the form 
\begin{equation}
[ \hat{x}^j, \: \hat{p}_5^k] = i \delta_{jk}  \gamma_5,
\end{equation}
which differs by $\gamma_5$ from the commutator  $[ \hat{x}^j, \: \hat{p}^k]$. In spite of the difference, 
 the implied uncertainty relation has  the usual form 
\[\langle \psi| (\Delta \hat{ x}^j)^2|\psi\rangle \langle \psi |(\Delta \hat{p}_5^k)^2|\psi\rangle \geq  \frac{1}{4} \delta_{jk}, \]  
where  $\Delta \hat{x}^j= \hat{x}^j - \langle\psi| \hat{x}^j|\psi \rangle$, $\;\Delta \hat{p}^k_5 = \hat{p}^k_5 - \langle\psi| \hat{p}^k_5|\psi \rangle$. The uncertainty relation is obtained in the standard manner.  Let us consider  
\[  I(\alpha) =  \langle \psi|  (\alpha \Delta \hat{p}^k_5  + i \gamma_5 \Delta \hat{x}^j)  (\alpha \Delta \hat{p}^k_5  -  i \gamma_5 \Delta \hat{x}^j)|\psi \rangle, \]
where $\alpha$ is a real variable.  It is clear that $I(\alpha) \geq 0$. On the other hand, using  commutator (2)  we have
\[ I(\alpha) = \alpha^2 \langle \psi|  (\Delta \hat{p}_5^k)^2|\psi \rangle - \alpha \delta_{jk} + \langle \psi|  (\Delta \hat{x}^j)^2|\psi \rangle. \] We know that this  quadratic polynomial in $\alpha $ does  not have  two distinct real roots.  The uncertainty relation follows as the necessary and sufficient condition for this.

In the massive case ($m>0$) the axial momentum has  nontrivial  time evolution in the Heisenberg picture, because it does not commute with the Hamiltonian $\hat{h}$ shown below.  This aspect is discussed in detail  in \cite{10}. 

\section{Time evolution of the axial  momentum amplitudes and quaternions}
The general solution of Eq.\ (1)  in the basis of  eigenfunctions of the axial momentum was found in \cite{10}.  It is  complete from theoretical viewpoint, but rather clumsy 
if one thinks about concrete applications.  Below we transform that solution  to a much simpler form.    Furthermore, we point out that  the general solution  can be described in terms of quaternions. Such a link of the Majorana quantum mechanics with the algebra of quaternions  is yet another intriguing feature of it,  in addition to the  non Hermitian Hamiltonian $\hat{h}$ and non conservation of the axial momentum in the case of free massive Majorana particle.  

Let us begin by recalling necessary facts from the paper \cite{10}.  The Dirac equation (1)  is rewritten as  
\begin{equation} 
\partial_t \psi = \hat{h} \psi, 
\end{equation}
where the Hamiltonian 
\[ \hat{h} = - \gamma^0 \gamma^k \partial_k - i m \gamma^0 \]
 is real and anti-symmetric, but it is not Hermitian if $m\neq0$.  Nevertheless,  the scalar product  $ \langle \psi_1(t)|\psi_2(t) \rangle = \int\!d^3x\: 
\psi_1^T(\mathbf{x}, t)\: \psi_2(\mathbf{x},  t)     $ is constant in time because  $\hat{h}$  is anti-symmetric.  The time evolution operator is orthogonal one.   

The normalized eigenfunctions of  the axial  momentum have the form 
\begin{equation}
 \psi_{ \mathbf{p}}(\mathbf{x}) =  (2\pi)^{-3/2} \exp(i \gamma_5 \mathbf{p}\mathbf{x}) \: v,    
\end{equation}
They  obey the conditions 
\[  \hat{\mathbf{p}}_5 \psi_{ \mathbf{p}}(\mathbf{x}) =  \mathbf{p}\:  \psi_{ \mathbf{p}}(\mathbf{x}),  \;\;\;\; \int\!d^3x\:  \psi^{T}_{ \mathbf{p}}(\mathbf{x})\: \psi_{\mathbf{q}}(\mathbf{x}) = \delta(\mathbf{p} - \mathbf{q}).  \] 
 where  $v$ is an arbitrary constant, normalized ($ v^T v=1$) and real  bispinor.   
We call the functions $\psi_{\mathbf{p}}(\mathbf{x})$ the axial plane waves \footnote{Note that the time dependence is not included  --  it is represented by a time-dependent $SO(4) $ matrix, see formulas (7) below.}.  Note that  \[  \exp(i \gamma_5 \mathbf{p}\mathbf{x}) =  \cos(\mathbf{p}\mathbf{x}) I  + i \gamma_5 \sin( \mathbf{p}  \mathbf{x} ). \]

The  expansion  of $\psi(\mathbf{x}, t) $  into the axial plane waves has the form 
\begin{equation}\psi(\mathbf{x}, t) = \frac{1}{(2\pi)^{3/2}} \sum_{\alpha=1}^2 \int\!d^3p\: e^{i \gamma_5 \mathbf{p} \mathbf{x}} \left(v_{\alpha}^{( +)}(\mathbf{p})  c_{\alpha}(\mathbf{p},t) +  v_{\alpha}^{( -)}(\mathbf{p})  d_{\alpha}(\mathbf{p},t)\right),  \end{equation}
 where the basis bispinors $v_{\alpha}^{(\pm)}$ obey the conditions  
\[ \gamma^0 \gamma^k p^k \:v^{(\pm)}_{\alpha} = \pm |\mathbf{p}| \:v^{(\pm)}_{\alpha}. \] 
 The eigenvalues $ \pm |\mathbf{p}|$ correspond to helicities $ \pm 1/2\;$, respectively, \cite{10}. They    are double degenerate ($\alpha=1,2$).   Thus each single mode in (5) is common normalized eigenstate of $\hat{\mathbf{p}}_5$ and of the helicity. 
The index $\alpha= 1,2$ reflects the degeneracy of the common eigenstates which is an artefact of the reality of our Hilbert space.

 The  basis  bispinors have the following form 
\begin{equation*}
v_1^{(+)}(\mathbf{p}) = \frac{1}{\sqrt{2  |\mathbf{p}|  (  |\mathbf{p}|  - p^2) }} \left(\begin{array}{c} - p^3\\  p^2 - |\mathbf{p}| \\ p^1 \\ 0   \end{array}  \right),    \;\;\; v_2^{(+)}(\mathbf{p}) =  i \gamma_5\: v_1^{(+)}(\mathbf{p}), 
\end{equation*}
\begin{equation}  
v_1^{(-)}(\mathbf{p}) =  i \gamma^0\: v_1^{(+)}(\mathbf{p}),  \;\;\; v_2^{(-)}(\mathbf{p}) =  i \gamma_5\: v_1^{(-)}(\mathbf{p}) = - \gamma_5 \gamma^0 v_1^{(+)}(\mathbf{p}). 
\end{equation}
They are orthonormal,
\[  (v^{(\epsilon)}_{\alpha})^T(\mathbf{p}) \: v^{(\epsilon')}_{\beta}(\mathbf{p}) = \delta_{\epsilon \epsilon'} \delta_{\alpha \beta}, \] 
where $\epsilon, \epsilon' = +, -$  refer to the helicity,  and $\alpha, \beta = 1,2.$  The basis (6)  has quite remarkable properties:   it is real;  generated  from $v_1^{(+)}$ by the quaternions  which are  introduced below; and it does not depend on the mass $m$ -- it is scale invariant. 
  
Time dependence of the real amplitudes $ c_{\alpha}(\mathbf{p},t), \;  d_{\alpha}(\mathbf{p},t) $ in expansion (5) is found  by solving Eq.\ (3). To this end,  the amplitudes are split into the even and odd parts, 
\[ c_{\alpha}(\mathbf{p},t) =  c_{\alpha}^{\:'}(\mathbf{p},t) +  c_{\alpha}^{\:''}(\mathbf{p},t),  \;\;\; d_{\alpha}(\mathbf{p},t) =  d_{\alpha}^{\:'}(\mathbf{p},t) +  d_{\alpha}^{\:''}(\mathbf{p},t),   \]
where $c_{\alpha}^{\:'}(- \mathbf{p},t) = c_{\alpha}^{\:'}(\mathbf{p},t), $ $\;c_{\alpha}^{\:''}(- \mathbf{p},t) =- c_{\alpha}^{\:''}(\mathbf{p},t)$, and analogously for $d', d''$. Furthermore, we  introduce the notation    
\[ \vec{\:c}(\mathbf{p},t) =  \left(\begin{array}{c} c_1' \\ c_1''\\ c_2'\\ c_2'' \end{array} \right)\!\!, \:
 \vec{\:d}(\mathbf{p},t) =  \left(\begin{array}{c} d_1' \\ d_1''\\ d_2'\\ d_2'' \end{array} \right)\!\!,\:  K_{\pm}(\mathbf{p}) = \left(\begin{array}{cccc} 0 & -n^1& \pm n^2&\pm n^3\\ n^1& 0 &\mp n^3 &\pm n^2\\ \mp n^2& \pm n^3 &0& n^1 \\ \mp n^3 & \mp n^2 & - n^1 &0 \end{array}\right)\!\!,  \]
where
\[  n^1= \frac{m\: p^1}{E_p \:\sqrt{(p^1)^2 + (p^3)^2}}, \;\; n^2 =  \frac{|\mathbf{p}|}{E_p}, \;\;  n^3= \frac{m \: p^3}{E_p\: \sqrt{(p^1)^2 + (p^3)^2}}, \] and  $ E_p=\sqrt{m^2 + \mathbf{p}^2}$.   
The time dependence of the amplitudes is  given by following formulas \cite{10}
\begin{equation} \vec{\:c}(\mathbf{p},t) =  \exp(t E_p \: K_+(\mathbf{p}))     \vec{\:c}(\mathbf{p},0),   \;\; \vec{\:d}(\mathbf{p},t) = \exp(t E_p\: K_-(\mathbf{p}))     \vec{\:d}(\mathbf{p},0).    \end{equation}
The  matrices $K_{\pm}(\mathbf{p})$ are anti-symmetric,  hence the matrices $ \exp(t E_p\: K_{\pm}(\mathbf{p}))$ belong to the $SO(4)$ group.   Because $K_{\pm}^2 = - I$,  we  have the formula
\begin{equation} \exp(t E_p\: K_{\pm}(\mathbf{p})) =  \cos(t E_p) I + \sin(t E_p) K_{\pm}(\mathbf{p}).  \end{equation}
Here we end the recapitulation of the relevant for this work facts from \cite{10}.

Using  expansion (5)  we immediately obtain the Plancherel formula
\[ \langle \psi_1| \psi_2 \rangle =  \int\! d^3p \: \sum^2_{\alpha=1} \left(c^{(1)}_{\alpha}(\vec{p},t) c^{(2)}_{\alpha}(\vec{p},t) + d^{(1)}_{\alpha}(\vec{p},t) d^{(2)}_{\alpha}(\vec{p},t)\right), \] 
where the amplitudes $c^{(1)}_{\alpha}, \:d^{(1)}_{\alpha}$ correspond to $\psi_1$, and    $c^{(2)}_{\alpha}, \: d^{(2)}_{\alpha} $ to $\psi_2$.  Let us remind that the integration variable $\mathbf{p}$ is the eigenvalue of the axial, not the ordinary,  momentum.    Splitting the amplitudes into the even and odd parts, we may write
\begin{equation} \langle \psi_1| \psi_2 \rangle =  \int\! d^3p \: \left( (\vec{c}^{\:(1)})^T \vec{c}^{\:(2)} +  (\vec{d}^{\:(1)})^T \vec{d}^{\:(2)} \right).   \end{equation}
Because the time evolution is given by the  orthogonal matrices, as shown in  formulas (7),  we again see that the scalar product is constant in time.

In the   general solution of Eq. (3)  quoted above  the amplitudes $c_{\alpha},\; d_{\alpha}$ are split into the even and odd components  which  mix  during the  time evolution,  see formulas (7).  It turns out that  the solution can be rewritten in a  more transparent form.  To this end,  we use formulas (7) and  (8), as well as  the concrete form (6)  of the basis bispinors. After straightforward and somewhat lengthy calculation the general solution  is transformed to the following form
\begin{equation}
\psi(\mathbf{x}, t) = \frac{1}{(2\pi)^{3/2}} \sum_{\alpha=1}^2 \int\!d^3p\: \left[\left(\cos(E_pt) I - \hat{L}_+  \sin(E_pt)  \right)   e^{i \gamma_5 \mathbf{p} \mathbf{x}} v_{\alpha}^{( +)}(\mathbf{p})  c_{\alpha}(\mathbf{p},0)  \right. \end{equation} \[ \;\;\;\;\;\;\;\;\;\;\left. +  \left(\cos(E_pt) I + \hat{L}_- \sin(E_pt)  \right)
  e^{i \gamma_5 \mathbf{p} \mathbf{x}}
 v_{\alpha}^{( -)}(\mathbf{p})  d_{\alpha}(\mathbf{p},0)\right],
\]
where 
\[ \hat{L}_{\pm} =  i \gamma_5 \frac{|\mathbf{p}|}{E_p}  \pm  i \gamma^0 \frac{m}{E_p}. \]
In  formula (10)  we have the initial values of  the full amplitudes $c_{\alpha}, \: d_{\alpha}$,  while in (7) the even and odd  parts appear separately.

Notice that 
\[ \cos(E_pt) \mp \hat{L}_{\pm}  \sin(E_pt) = \exp(\mp \hat{L}_{\pm} E_p t), \] because  $\hat{L}_{\pm}^2 = -I$. Therefore, 
in the massless case, i.e.,  $m=0$, formula (10) acquires a very simple form, namely  
\begin{equation}
\psi(\mathbf{x}, t) = \frac{1}{(2\pi)^{3/2}} \sum_{\alpha=1}^2 \int\!d^3p\: \left[  e^{i \gamma_5 (\mathbf{p} \mathbf{x}  - |\mathbf{p}| t)} v_{\alpha}^{( +)}(\mathbf{p}) c_{\alpha}(\mathbf{p},0)   \right.  \end{equation} \[ \;\;\;\;\;\;\;\;\; \;\;\;\;\;\;\;\;\;  \left.+  e^{i \gamma_5 (\mathbf{p} \mathbf{x}  + |\mathbf{p}| t)}v_{\alpha}^{( -)}(\mathbf{p})  d_{\alpha}(\mathbf{p},0)  \right]
\]
Thus, in this case the modes with different helicity do not mix during time evolution, in accordance with the theory of  irreducible representations of the Poincar\'e group.

Intriguingly, the general solution (7)  and its equivalent form (10)  can be rewritten in terms of quaternions. 
The quaternionic units $\hat{i}, \hat{j}, \hat{k}$ are introduced as follows:
\[ \hat{i}= i \gamma_5, \;\;\;   \hat{j}= i \gamma^0, \;\;\;   \hat{k}= - \gamma_5 \gamma^0 = i \gamma^1 \gamma^2 \gamma^3.\]
They obey the usual conditions
\[ \hat{i}^2 = \hat{j}^2 = \hat{k}^2 = - I, \;\;\; \hat{i}\hat{j}=\hat{k}, \;\;\; \hat{k}\hat{i}=\hat{j}, \;\;\; \hat{j}\hat{k}=\hat{i}. \]
The bispinor basis $v^{(\pm)}_{\alpha}(\mathbf{p})$ is generated from $v^{(+)}_1(\mathbf{p})$ by acting with $\hat{i}, \hat{j}, \hat{k}$, see formulas (6).   Moreover, 
all matrices present in formulas (7) and (10) can be expressed by $I, \hat{i}, \hat{j}, \hat{k}$. In  particular,  $K_{\pm}(\mathbf{p}) =   \mp n^2 \hat{i} \pm n^3 \hat{j} +  n^1 \hat{k}$. Therefore,  the time evolution of the amplitudes $\vec{c}, \: \vec{d}$ at each fixed value of the axial momentum $\mathbf{p}$  is given by a time dependent quaternion.  

Solution (10) can be written in a Fourier form, in which no matrices are present,  only trigonometric functions and the basis bispinors  (6).  This is possible because the quaternions acting on the basis bispinors  do not yield any  new bispinors, but only  permute them.  This form of  solution (10)  reads 
\begin{equation*} \psi(\mathbf{x}, t) = \frac{1}{(2\pi)^{3/2}} \int\!d^3p\: \left[\cos(\mathbf{p}\mathbf{x}) \cos(E_pt) V_{cc}(\mathbf{p})  + \cos(\mathbf{p}\mathbf{x}) \sin(E_pt) V_{cs}(\mathbf{p}) \right. \end{equation*}  \begin{equation} \left.+ \sin(\mathbf{p}\mathbf{x}) \cos(E_pt) V_{sc}(\mathbf{p})+ \sin(\mathbf{p}\mathbf{x}) \sin(E_pt) V_{ss}(\mathbf{p})   \right],
\end{equation}
where the $V_{..}$ stand  for linear combinations of the basis  bispinors, namely
\[ V_{cc}(\mathbf{p})= c_1(\mathbf{p},0) v^{(+)}_1(\mathbf{p}) + c_2(\mathbf{p},0) v^{(+)}_2(\mathbf{p}) +d_1(\mathbf{p},0) v^{(-)}_1(\mathbf{p})+d_2(\mathbf{p},0) v^{(-)}_2(\mathbf{p}), \]
\[V_{cs}(\mathbf{p}) =\frac{1}{E_p} \left[ \left( m\: d_1(\mathbf{p},0) + |\mathbf{p}| c_2(\mathbf{p},0)     \right) v^{(+)}_1(\mathbf{p})  \right.  \;\;\;\;\;\;\;\;\;\;\;\;\;\; \;\;\;\;\;\;\;\;\;\;\;\;\;\; \;\;\;\;\;\;\;\;\;\;\;\;\;\;\;\;\;   \]  \[  -  \left( m \:d_2(\mathbf{p},0) + |\mathbf{p}| c_1(\mathbf{p},0)     \right) v^{(+)}_2(\mathbf{p}) 
  - \left( m\: c_1(\mathbf{p},0) + |\mathbf{p}| d_2(\mathbf{p},0)     \right) v^{(-)}_1(\mathbf{p})  \]  \[  \left.  \;\;\;\;\;\;\;\;\;\;\;\;\;\; \;\;\;\;\;\;\;\;\;\;\;\;\;\; \;\;\;\;\;\;\;\;\;\;\;\;\;\;  + \left( m\: c_2(\mathbf{p},0) + |\mathbf{p}| d_1(\mathbf{p},0)     \right) v^{(-)}_2(\mathbf{p})   \right], \]
\[V_{sc}(\mathbf{p})=- c_2(\mathbf{p},0) v^{(+)}_1(\mathbf{p}) + c_1(\mathbf{p},0) v^{(+)}_2(\mathbf{p}) - d_2(\mathbf{p},0) v^{(-)}_1(\mathbf{p})+d_1(\mathbf{p},0) v^{(-)}_2(\mathbf{p}), \]   
\[V_{ss}(\mathbf{p}) =\frac{1}{E_p} \left[- \left( m\: d_2(\mathbf{p},0) - |\mathbf{p}| c_1(\mathbf{p},0)     \right) v^{(+)}_1(\mathbf{p})  \;\;\;\;\;\;\;\;\;\;\;\;\;\; \;\;\;\;\;\;\;\;\;\;\;\;\;\;\;\;\;\;\;\;\;\;\;\;\;\;\; \right. \]  \[ -  \left( m\: d_1(\mathbf{p},0) - |\mathbf{p}| c_2(\mathbf{p},0)     \right) v^{(+)}_2(\mathbf{p}) 
 + \left( m\: c_2(\mathbf{p},0) - |\mathbf{p}| d_1(\mathbf{p},0)     \right) v^{(-)}_1(\mathbf{p}) \]   \[  \left.  \;\;\;\;\;\;\;\;\;\;\;\;\;\; \;\;\;\;\;\;\;\;\;\;\;\;\;\; \;\;\;\;\;\;\;\;\;\;\;\;\;\; +  \left( m\: c_1(\mathbf{p},0) - |\mathbf{p}| d_2(\mathbf{p},0)     \right) v^{(-)}_2(\mathbf{p})   \right]. \]
Formula  (12)  is a convenient starting point for analysis of concrete examples of solutions. 

Solution (12)  is a superposition of standing plane waves.  In order to rewrite it in terms of traveling plane waves we 
use  trigonometric formulas such as $   \cos(\mathbf{p}\mathbf{x}) \cos(E_pt) = \frac{1}{2} (\cos(\mathbf{p}\mathbf{x} - E_p t) + \cos(\mathbf{p}\mathbf{x} + E_p t))$, etc.  We  obtain
\begin{equation}
\psi(\mathbf{x}, t) = \frac{1}{2(2\pi)^{3/2}} \int\!d^3p\: \left[ \cos(\mathbf{p}\mathbf{x} - E_pt) \: A_+(\mathbf{p})  + \cos(\mathbf{p}\mathbf{x} + E_pt) \:A_-(\mathbf{p})  \right. 
\end{equation} 
\[  \left.   \;\;\;\;\;\;\;\;\;\;\;\;\;\; \;\;\;\;\;\;\;\;\;\;\;\;\;\; +\sin(\mathbf{p}\mathbf{x} - E_pt) \: B_+(\mathbf{p})  + \sin(\mathbf{p}\mathbf{x} + E_pt) \:B_-(\mathbf{p}) \right], \] 
where 
\[ A_{\pm}(\mathbf{p}) = v^{(+)}_1(\mathbf{p}) A^{1}_{\pm}(\mathbf{p})  +v^{(+)}_2(\mathbf{p}) A^{2}_{\pm}(\mathbf{p}) +v^{(-)}_1(\mathbf{p}) A^{3}_{\pm}(\mathbf{p}) +v^{(-)}_2(\mathbf{p}) A^{4}_{\pm}(\mathbf{p}),   \]  
\[ B_{\pm}(\mathbf{p}) = v^{(+)}_1(\mathbf{p}) B^{1}_{\pm}(\mathbf{p})  +v^{(+)}_2(\mathbf{p}) B^{2}_{\pm}(\mathbf{p}) +v^{(-)}_1(\mathbf{p}) B^{3}_{\pm}(\mathbf{p}) +v^{(-)}_2(\mathbf{p}) B^{4}_{\pm}(\mathbf{p}),   \]  
and 
\[ A^{1}_{\pm} =   (1 \pm \frac{p}{E_p})  c_1   \mp   \frac{m}{E_p} d_2,  \;\;\; A^{2}_{\pm} =   (1 \pm \frac{p}{E_p}) c_2  \mp  \frac{m}{E_p} d_1, \] 
\[  A^{3}_{\pm} =   (1 \mp \frac{p}{E_p}) d_1   \pm   \frac{m}{E_p} c_2, \;\;\;  A^{4}_{\pm} =  (1 \mp \frac{p}{E_p}) d_2   \pm   \frac{m}{E_p} c_1, \]
\[ B^{1}_{\pm} =  -  (1 \pm \frac{p}{E_p})  c_2   \mp   \frac{m}{E_p} d_1,  \;\;\; B^{2}_{\pm} =   (1 \pm \frac{p}{E_p}) c_1  \pm  \frac{m}{E_p} d_2, \] 
\[  B^{3}_{\pm} = -  (1 \mp \frac{p}{E_p}) d_2   \pm   \frac{m}{E_p} c_1, \;\;\;  B^{4}_{\pm} =  (1 \mp \frac{p}{E_p}) d_1   \mp   \frac{m}{E_p} c_2. \]
In these formulas $p \equiv |\mathbf{p}|$,  $E_p = \sqrt{\mathbf{p}^2+ m^2}, $  and the amplitudes  $c_1,  c_2, d_{1},  d_2$ are the ones present in formula (10) (the arguments $(\mathbf{p}, 0)$ have been omitted for brevity).  Let us remind again that $\mathbf{p}$ is the eigenvalue of the axial momentum.

Let us  consider now  a single mode with fixed value $\mathbf{q}$ of the axial  momentum, i.e., we put in the formulas above $c_{\alpha}(\mathbf{p},0) = c_{\alpha}\: \delta(\mathbf{p}- \mathbf{q}), \:d_{\alpha}(\mathbf{p},0) =  d_{\alpha}\: \delta(\mathbf{p}- \mathbf{q})$, where $c_{\alpha}, d_{\alpha}$, $\alpha=1,2,$ are constants now.  In the massless case,
\[ A^1_+ = 2c_1, \; A^2_+ = 2c_2, \;  A^3_+= A^4_+=  A^1_- =A^2_- =0, \;A^3_- = 2d_1, \; A^4_- = 2d_2, \] 
\[ B^1_+ =- 2c_2, \; B^2_+ = 2c_1, \;  B^3_+= B^4_+=  B^1_- =B^2_- =0, \;B^3_- =-2d_2, \; B^4_- = 2d_1. \] 
We see that in this case the $A_+, B_+$ part   on the r.h.s. of formula (13)  is independent of the  $A_-, B_-$ part. In particular,   we can put one of them to zero in order  to obtain a plane wave propagating in the direction of $\mathbf{q}$  or $-\mathbf{q}$.  

The massive case is different -- the plane wave always has the two components propagating in the opposite directions, $\mathbf{q}$ and $-\mathbf{q}$.  If we assume that $A_-=0$,  simple calculation shows that also $ B_- =A_+=B_+=0$;   if we put $B_- =0$ then   
also $ A_- =A_+=B_+=0$.

Let  us put  the constants $d_1 =d_2=0$.  In the massless case this assumption   gives the plane wave moving in the direction $\mathbf{q}$,
\[ \psi(\mathbf{x}, t) = \frac{1}{(2\pi)^{3/2}}   \cos(\mathbf{q}\mathbf{x} - E_qt) \: (c_1 v^{(+)}_1(\mathbf{q})  +c_2 v^{(+)}_2(\mathbf{q}) )  \;\;\;\;\;\;\;\;\;\;\; \;\;\;\;\;\;\;\;\;\;\; \]   \[  \;\;\;\;\;\;\;\;\;\;\; \;\;\;\;\;\;\;\;\;\;\; + \frac{1}{(2\pi)^{3/2}}\sin(\mathbf{q}\mathbf{x} - E_qt) \:(-  c_2 v^{(+)}_1(\mathbf{q})  +c_1 v^{(+)}_2(\mathbf{q})   ). \] 

  In the massive case  all four components in (13) do not vanish. 
However,   the amplitudes of the $- \mathbf{q}$  components, i.e., $A_-, B_-  $ ,  are negligibly small in the high frequency limit, i.e., when  $m/E_q \ll 1$. In this limit  
\[ A^1_+ \approx 2 c_1, \; A^2_+ \approx 2 c_2, \;  A^3_+ =\frac{m}{E_q} c_2 , \; A^4_+ = \frac{m}{E_q} c_1,  \]  \[ B^1_+ \approx -2 c_2, \; B^2_+ \approx 2 c_1, \;  B^3_+ =\frac{m}{E_q} c_1 , \; B^4_+ = - \frac{m}{E_q} c_2,  \]
and \[  A^1_- \approx \frac{m^2}{2 E_q^2} c_1, \;A^2_- \approx \frac{m^2}{2E_q^2} c_2, \;A^3_- = -\frac{m}{E_q}c_2, \; A^4_- = - \frac{m}{E_q}c_1, \] 
\[  B^1_- \approx -\frac{m^2}{2 E_q^2} c_2, \;B^2_- \approx \frac{m^2}{2E_q^2} c_1, \;B^3_- = -\frac{m}{E_q}c_1, \; B^4_- = \frac{m}{E_q}c_2. \] 
On the other hand, in the limit of long waves, i.e., $q \ll m$, 
\[A^1_{\pm} \approx c_1, \; A^2_{\pm} \approx  c_2, \;  A^3_{\pm} \approx \pm c_2 , \; A^4_{\pm} \approx \pm  c_1 ,   \]   and  \[ B^1_{\pm} \approx -c_2, \; B^2_{\pm} \approx  c_1, \;  B^3_{\pm} \approx  \pm c_1, \; B^4_= \approx  \mp c_2.  \] 
In this case  the  $\mathbf{q}$ and $-\mathbf{q}$ components  have approximately equal  magnitudes. 

\section{Relation with  irreducible representations of the Poincar\'e group} 
 The Poincar\'e  transformations of the real bispinor $\psi(x)$ have the standard form,  \begin{equation} \psi'(x) = S(L) \psi(L^{-1}(x-a)), \end{equation}
with  $S(L) = \exp(\omega_{\mu\nu} [\gamma^{\mu}, \gamma^{\nu}]/8)$, where $\omega_{\mu\nu}= - \omega_{\nu\mu}$ parameterize the proper orthochronous Lorentz group,  $L=\exp(\omega^{\mu\;\;}_{\;\;\nu})$, in a vicinity of the unit element.  Below we show that  in the massive case these transformations imply transformations of the axial momentum dependent amplitudes which coincide with the real form of a single standard unitary Wigner's representation with spin 1/2. This representation being real and unitary  is in fact  orthogonal one. The conclusion is that as far as the relativistic transformations is the issue,  the expansion into the axial plane waves of the Majorana bispinor has the properties  expected for a single massive spin 1/2 particle.

 We do not discuss  here   representations of the Poincar\'e group pertaining to  the  massless  Majorana particle ($m=0$). There are two independent irreducible representations with the helicities $\pm 1/2$. This case is simpler because now the operator $\hat{\mathbf{p}}_5$  commutes with the Hamiltonian.  It will be presented in  pedagogical notes  \cite{14}.

We start from the following  expansion  into the eigenstates of the axial momentum
\begin{equation}
\psi(\mathbf{x}, t) =\frac{1}{ (2\pi)^{3/2}}  \int\! \frac{d^3p}{E_p}\: e^{i \gamma_5 \mathbf{p} \mathbf{x}} \:   v(\mathbf{p}, t), 
 \end{equation}
where $v(\mathbf{p}, t)$  is a real bispinor, and   $E_p = \sqrt{m^2 + \mathbf{p}^2}$.   Equation (3)  gives time evolution equation for $v$  
\begin{equation}
\dot{v}(\mathbf{p}, t)  = - i \gamma^0 \gamma^k \gamma_5 p^k v(\mathbf{p}, t) - im \gamma^0 v(-\mathbf{p}, t).  
\end{equation}
The reason for $v(-\mathbf{p}, t)$   in the last term on the r.h.s.  is that $\gamma^0$ anti-commutes with $\gamma_5$ and therefore
$\gamma^0 \exp(i \gamma_5 \mathbf{p} \mathbf{x}) =   \exp(- i \gamma_5 \mathbf{p} \mathbf{x}) \gamma^0.$
Taking time derivative of  Eq.\ (16)  we obtain the following equation
\[ \ddot{v}(\mathbf{p}, t) = - E_p^2 v(\mathbf{p}, t). \]  Let us write its general solution in the form 
\begin{equation}
v(\mathbf{p}, t) = \exp(-i\gamma_5 E_p t) v_+(\mathbf{p})  +   \exp(i\gamma_5 E_p t) v_-(-\mathbf{p}), 
\end{equation}
where the argument  of $v_-$ is  $-\mathbf{p}$  for later convenience. 
Then $\psi(\mathbf{x}, t) $ can be written as
\begin{equation}
\psi(\mathbf{x}, t) =\frac{1}{ (2\pi)^{3/2}}  \int\! \frac{d^3p}{E_p}\:\left( e^{i \gamma_5 (\mathbf{p} \mathbf{x}-E_pt)}\:   v_+(\mathbf{p})  +   e^{-i \gamma_5 (\mathbf{p} \mathbf{x}-E_pt)}\:   v_-(\mathbf{p})   \right),
 \end{equation}
where in the last term we have changed the integration variable to $- \mathbf{p}$. 
Furthermore,  Eq.\ (16)  is satisfied   by $v(\mathbf{p}, t)$ of the form (17) only if   $ v_{\pm}(\mathbf{p})$ obey the following conditions
\begin{equation}
E_p  \gamma_5 v_{\pm}(\mathbf{p}) = \gamma^0 \gamma^k p^k \gamma_5 v_{\pm}(\mathbf{p}) \pm m \gamma^0  v_{\mp}(\mathbf{p}) . 
\end{equation}

Applying the transformation law (14) with $a=0$  to  solution  (18), we obtain  Lorentz transformation of the bispinors $ v_{\pm}(\mathbf{p})$,    
\begin{equation}
v_{\pm}^{'}(p) =   S(L) \:v_{\pm}(L^{-1}p),
\end{equation}
where now we use the four-vector $p$ instead of $\mathbf{p}$  for  convenience in notation:   $v_+(p) \equiv v_+(\mathbf{p})$  and $p^0 =E_p$.  
The spacetime  translations $x'  = x + a  $  are represented by $SO(4)$    factor 
\begin{equation}  v'_{\pm}(\mathbf{p}) = e^{\pm i \gamma_5  p a}   v_{\pm}(\mathbf{p}). \end{equation}

 In the massive  case,  $v_-(\mathbf{p}) $  can be expressed by $v_+(\mathbf{p})$, see (19). The scalar product $\langle\psi_1|\psi_2\rangle = \int \!d^3x\:\psi_1^T(\mathbf{x},t) \psi_2(\mathbf{x}, t) $ acquires  explicitly Poincar\'e  invariant (and time independent) form 
\begin{equation} \langle \psi_1 | \psi_2 \rangle = \frac{2}{m^2}  \int\!\frac{d^3p}{E_p}\: \overline{v_{1+}(\mathbf{p})}\: ( \gamma^0 E_p - \gamma^k p^k)\: v_{2+}(\mathbf{p}),\end{equation} 
where $\overline{v_{1+}(\mathbf{p})} = v_{1+}^T(\mathbf{p}) \gamma^0$,  and $v_{1+}$ $(v_{2+}) $ corresponds to $\psi_1$ $(\psi_2)$ by formula (18). 
 
Transformations  (20), (21)  are unitary with respect to this scalar product. Thus, we have here  real unitary, i.e., orthogonal,  representation of the Poincar\'e  group.  It  turns out that it is irreducible and equivalent to a real version of the standard spin 1/2 unitary representation.  Detailed analysis 
of the representation is given  below.   Let us recall that   in the case of  massive Dirac particle  one finds a reducible representation which is  a direct sum of   two  spin 1/2 representations. 

Representation (20) can be cast in the standard  form which involves the Wigner rotations and a representation of  $SU(2)$ group \cite{15}.
To this end,  we introduce the standard momentum $\stackrel{(0)}{p} = (m, 0,0,0)$, where $m >0$,  as well as a Lorentz boost $H(p)$ such that  $H(p) \stackrel{(0)}{p} = p$.  Furthermore, at each $p$ we introduce the basis  of real bispinors,
\begin{equation} v_i(p) = S(H(p)) v_i(\stackrel{(0)}{p}),\end{equation} 
  where $i = 1, 2, 3, 4,$ and $ v_i(\stackrel{(0)}{p})$ is a basis at $ \stackrel{(0)}{p}$ such that  $m v^T_i(\stackrel{(0)}{p})  v_k(\stackrel{(0)}{p}) = \delta_{ik}$ (the factor $m$ is included for dimensional reason). 
Here again we use the four momentum in the  notation as in (20).  The bispinor $v_+(p)$  
is decomposed in this basis,  \[v_+(p) = a^i(p) v_i(p).\]  The scalar product (22)  is equal to \begin{equation}  \langle \psi_1| \psi_2 \rangle = \frac{2}{m^2} \int \! \frac{d^3 p}{E_p} \: a_1^k(p) a_2^k(p), \end{equation}
where  $k= 1,2,3,4$.  The real dimensionless amplitudes $a^k_1, a^k_2$  correspond to $\psi_1, \psi_2$, respectively. 

Let us find the relativistic transformation law of the amplitudes $a^k(p)$. In the case of Lorentz transformations,  using (20)  we have 
\[  v'_+(p) = a^{'k}(p) v_k(p) = S(L) a^i(L^{-1}p) v_i(L^{-1}p)  \;\;\;\;\;\; \]  \[    =  a^i(L^{-1}p) S(H(p)) S(H^{-1}(p) L H(L^{-1}p)) v_i( \stackrel{(0)}{p} ).    \] 
The Lorentz transformation ${\cal R}(L,p)) =H^{-1}(p) L H(L^{-1}p)$  leaves  $\stackrel{(0)}{p}$  invariant -- it is a rotation, known as the Wigner rotation.  Therefore, we may write 
\begin{equation}
S({\cal R}(L,p)) v_i( \stackrel{(0)}{p} ) =  D_{ki}({\cal R}(L,p))  v_k( \stackrel{(0)}{p} ). \end{equation}
  In consequence,  \[v'_+(p) = a^i(L^{-1}p) D_{ki}({\cal R}(L,p)) v_k(p),   \] and  finally
\begin{equation}
a^{'k}(p) =  D_{ki}({\cal R}(L,p))  a^i(L^{-1}p).
\end{equation}
The  invariance of the scalar product  (24)   implies orthogonality of the 4 by 4 real matrix  with the elements $D_{ki}({\cal R}(L,p))$.

The space-time translation  $\psi'(x) = \psi(x-a)$  results in a change of the amplitudes \footnote{We use the notation $\underline{a}$  because $a'$ is already  used.}, $ a^i(p) \rightarrow \underline{a}^i(p).$  Let us compute  $\underline{a}^i(p)$.  Using formula (21) we have 
\[ v'_+(p) = \underline{ a}^k(p) v_k(p) = e^{i \gamma_5 pa} a^k(p) v_k(p) = a^k(p) S(H(p))  e^{i \gamma_5 pa} v_k( \stackrel{(0)}{p} ) \]
\[  \hspace*{4cm}   = m a^k(p) S(H(p)) \left(v_l^T( \stackrel{(0)}{p} ) e^{i \gamma_5 pa}  v_k( \stackrel{(0)}{p} ) \right)  v_l( \stackrel{(0)}{p} ). \]
At this point it is convenient to choose the  basis   $v_k( \stackrel{(0)}{p} )$  in the Kronecker form, in which the $i$-th component of the bispinor  $ v_k( \stackrel{(0)}{p} )$  is equal to $\delta_{ik}/\sqrt{m}$.  In this basis 
\[ m  v_l^T( \stackrel{(0)}{p} ) e^{i \gamma_5 pa}  v_k( \stackrel{(0)}{p} )  = ( e^{i \gamma_5 pa})_{lk}.\]  In consequence, 
\begin{equation}
\underline{a}^l(p) =  ( e^{i \gamma_5 pa})_{lk} a^k(p).   
\end{equation}
The matrix  $ e^{i \gamma_5 pa}$ is  orthogonal, and the scalar product  (24)  is of course invariant  with respect to  the transformations (27).

Formula (25) opens the way to identification of the pertinent  orthogonal  representation of the Poincar\'e group.  This representation  is uniquely characterized by representation (26)  of the Wigner rotations \cite{15}. In order to identify this last representation it suffices to take  in formula (25)  $p =  \stackrel{(0)}{p}$ and $ L =R$, where $R$ is arbitrary rotation.  Then ${\cal R}(R,  \stackrel{(0)}{p}) = R$.  Let us again use the Kronecker basis introduced above. In this basis, formula (25) can now be rewritten as  $ S(R)= D(R)$, where the matrix elements of $D(R)$  are equal to   $D_{ki}(R)$.  Therefore we now turn to the matrices $S(R)$. 

The matrices $S(R)$ have the form 
\[  S(R) =  \exp(\frac{1}{2} (\omega_{12} \gamma^1\gamma^2 + \omega_{31} \gamma^3\gamma^1 +  \omega_{23} \gamma^2\gamma^3)).  \]
They form a subgroup of the SO(4) group. 
There exist real orthogonal  matrices ${\cal O}$ such that 
\[ {\cal O} \gamma^1 \gamma^2 {\cal O}^{-1} = \hat{i}, \;\;    {\cal O} \gamma^2 \gamma^3 {\cal O}^{-1} = \hat{j}, \;\;  {\cal O} \gamma^3 \gamma^1 {\cal O}^{-1} = \hat{k},   \]   
where $\hat{i}, \hat{j}, \hat{k}$ are the quaternions introduced in the previous section. For example, one may take  the matrix
\[{\cal O} = \frac{1}{\sqrt{2}}  \left(\begin{array}{cc} \sigma_0 & - \sigma_1\\
- \sigma_0 & -\sigma_1 \end{array} \right). \]
Thus, the matrices  ${\cal O} S(R) {\cal O}^{-1}$  are elements of the algebra of quaternions, and as such they can be written  in the form
\begin{equation} {\cal O} S(R) {\cal O}^{-1} = s_0 I_4 + s_1 \hat{i} + s_2 \hat{j} + s_3 \hat{k}, \end{equation}
where $s_0, s_k$ are real functions of the   parameters $\omega_{ik}$.  These matrices also belong to the $SO(4)$ group. Furthermore,  because
\[ ({\cal O} S(R) {\cal O}^{-1})^T = s_0 I_4 - s_1 \hat{i} -  s_2 \hat{j}  - s_3 \hat{k}\] 
and  ${\cal O} S(R) {\cal O}^{-1}  ({\cal O} S(R) {\cal O}^{-1})^{T} = I_4,$
we obtain the  relation  $(s_0)^2 + (s_1)^2 + (s_2)^2 + (s_3)^2 =1$.

On the other hand,  let us consider the spin 1/2  representation  $T(u)$ of $SU(2)$ group,  $T(u) \xi = u \xi$,  where $u\in SU(2)$ and $\xi$ is  a two-component spinor (in general complex).  This representation can be rewritten in  real form  simply by using  the real and imaginary parts.  Thus  we write  
\[  u = \left( \begin{array}{cc} \alpha & \ -\beta \\ \beta^* & \alpha^*   \end{array}   \right), \;\;\; \xi = \left( \begin{array}{c} \xi_1 \\   \xi_2 \end{array}  \right), \] 
where $\alpha= \alpha' + i \alpha'',  \; \beta = \beta' + i \beta'', \;  \xi_1= \xi_1' + i \xi_1'', \; \xi_2= \xi_2' + i \xi_2'', $ and 
$ \alpha \alpha^* + \beta \beta^* =(\alpha')^2 +(\alpha'')^2 + (\beta')^2 +(\beta'')^2 =1.$
Next, we form the four-component real vector $\vec{\xi}$ and the 4 by 4 real matrix $\hat{T}(u)$: \[
\vec{\xi} = \left(  \begin{array}{c} \xi_1' \\ \xi_1'' \\ \xi_2' \\ \xi_2''  \end{array}   \right), \;\;\; \hat{T}(u) = \left( \begin{array}{cccc} \alpha' & - \alpha'' & -\beta' &  \beta'' \\  \alpha'' & \alpha' & -\beta'' &  -\beta' \\  \beta' &  \beta'' & \alpha' & \alpha'' \\  -\beta''&  \beta' & -\alpha'' & \alpha' 
\end{array}  \right).  \]
It turns out that $\vec{\;\xi_u} = \hat{T}(u) \vec{\xi}, \;$ where $\xi_u \equiv T(u) \xi$. The matrix $\hat{T}(u)$ can be 
rewritten in terms of the quaternions, 
\begin{equation} \hat{T}(u) = \alpha' \:I_4 + \beta' \: \hat{i}+  \beta'' \: \hat{j}+  \alpha'' \:\hat{k}. \end{equation}
The r.h.s. of this formula  coincides with the r.h.s. of formula (28) if $  \alpha' = s_0,  \beta'  =s_1,   \beta'' = s_2$, and $ \alpha'' = s_3.   $  

In conclusion,   the representation of the Wigner rotations  given by the matrices $S(R)$  is equivalent to the real form of the spin 1/2 representation $T(u)$ of  $SU(2)$ group.  Thus, we have found that the representation of the Poincar\'e group  is the spin 1/2, $m>0$ representation. 
 Notice  that we have obtained  just one such spin 1/2 representation. In the  case of  Dirac particle a direct sum of two spin 1/2 representations  appears, one for particle and the other for antiparticle.

\section{Summary and remarks}
   1.  Let us summarize our main  results.  We have shown that the axial momentum operator for the Majorana particle is related to the ordinary momentum for the Weyl particle by the one-to-one mapping between the two models, and that it obeys the  Heisenberg
uncertainty relation. Next, using the eigenfunctions and eigenvalues of the axial momentum operator, we have written the general solution of the Dirac equation for the real bispinor in the form of superposition of traveling plane waves, with the eigenvalues $\mathbf{p}$ of the  axial momentum playing the role of wave vectors, i.e., giving the wave length and the direction of propagation.  In the massive case this superposition has the special feature that the plane waves  come in pairs with the opposite axial momenta,  $\mathbf{p}$ and $-\mathbf{p}$    This is a  consequence of the fact that in the massive case the axial momentum does not commute with the Hamltonian $\hat{h}$ .  Therefore, the eigenvectors of  $\hat{\mathbf{p}}_5$ are not stationary states -- the minimal stationary subspace in the Hilbert space is spanned by the two modes $\mathbf{p}, -\mathbf{p}$.  The presence of   such paired plane waves could perhaps serve  as a signature  of  the massive Majorana particle. This effect is relatively small at high energies,  but quite sizable at  energies close to the rest mass of the particle.     Finally,  we have shown that using the axial momentum basis one can unveil  the pertinent  irreducible spin 1/2 representation of the Poincar\'e group.

Apart from the results listed above, there are quite interesting purely theoretical aspects, namely  the  reformulation in terms of quaternions,  and  fully-fledged relativistic quantum mechanics over the algebraic field of real numbers $\mathbb{R}$ in place of complex numbers. 

     We conclude that  the axial momentum   $\hat{\mathbf{p}}_5 = -i \gamma_5 \nabla$ can be accepted as  the replacement for the ordinary momentum  $\hat{\mathbf{p}} = -i \nabla$. This latter operator  is not an observable for the Majorana particle  because it  does not commute with the charge conjugation $C$,  in contradistinction to $\hat{\mathbf{p}}_5$ .  In fact, we think that the axial momentum is 
the proper observable to be used in theoretical analysis of experimental data  for relativistic Majorana particles, when they are available.  \\

\noindent 2.   The  investigations of the axial momentum can be continued in several directions.  In our opinion, two are especially interesting. First,  we would like to check  time evolution of wave packets with certain fixed initial profile of the axial momentum.  Formulas (12) and (13) seem to be a good starting point for  work in this direction.  We believe that such a basic knowledge about evolution of wave functions more general than the plane waves can  facilitate searches for  the Majorana particles. 

The second  very interesting topic is the application of the axial plane waves  in quantum theory of the Majorana field. 
The amplitudes  $a^i(p)$ introduced right above formula (24),  where $p=(p^0, \mathbf{p})$ with  $\mathbf{p}$ being eigenvalue of the axial momentum and  $p^0 = \sqrt{m^2 + \mathbf{p}^2}$,   have clear transformation law with respect to the Poincar\'e  group.  This fact suggests that  precisely these amplitudes should be replaced by  creation and annihilation operators of the Majorana particle when quantizing the Majorana field. 

Finally, one may apply the axial momentum instead of the ordinary momentum in quantum mechanics of the Dirac particle.    Here the ordinary momentum    has the advantage -- it commutes  with the Dirac Hamiltonian  in the case of  free particle  --  but the use of  the axial momentum, which is after all  a legitimate observable,  can lead to new insights.

\section{Acknowledgement}
H. A. acknowledges a partial support from the Marian Smoluchowski Institute of Physics, Jagiellonian University, under Contract No.  337.1104.112.2019.

\end{document}